\documentstyle[pajasz,twoside,fleqn,epsf]{article} 
\begin{document}

\title{THEORETICAL STUDIES OF QUANTUM INTERFERENCE IN ELECTRONIC TRANSPORT
THROUGH CARBON NANOTUBES}

\author{W. IWO BABIACZYK and B. R. BU{\L}KA}

\address{Institute of Molecular Physics, Polish Academy of
  Sciences \\  Smoluchowskiego 17, 60-179 Pozna\'n, Poland}
\date{\today}

\maketitle                   
\begin{abstract}
 We performed studies of coherent electronic transport through a single walled carbon
 nanotube. In the calculations multiple scattering on the contacts and interference processes were
 taken into account. Conductance is a composition of contributions from different
 channels. We studied also spin--dependent transport in the system with ferromagnetic electrodes.
 The magnetoresistance is large and shows large oscillations, it can be even negative in some
 cases.
 \end{abstract}

\section{Introduction}

Recent experiments \cite{1,2,3} on electronic transport through
carbon nanotubes (CNT) show series of interesting effects.
Changing the length of the nanotube one can change a character of
the transport from the classical diffusive flow to the quantum
transport. The way of coupling of the CNT to the electrodes is
crucial for the transport properties. If the contact resistances
are large ($R\gg R_Q=$ 13 k$\Omega$), the transport exhibits the
incoherent  single-electron tunnelling character  with the
Coulomb blockade effect \cite{1}. Improvement of the quality of
the contacts results in an increase of the conductance of the
system and one can observe the Kondo resonance, which is due to
exchange interactions of conducting electrons with an
uncompensated spin localized on the CNT \cite{2}. In the case
where the contacts are made very carefully and thus the coupling
is strong, the transport through the system becomes coherent
\cite{3}. The interference processes and multiple scattering on
the contacts are very important in this limit and lead to the
Fabry-Perot interference.

We are interested in the coherent transport through  molecular
systems, particularly in magnetoresistance properties. Our system
is presented in figure 1, where the single walled carbon nanotube
(SWCNT) is strongly coupled to ferromagnetic electrodes (e.g. Ni).
Coherent spin-dependent transport
 measurements were already performed on multiwalled CNT \cite{4}.
 Despite  the small value of the resistance ($R\ge$9 k$\Omega$) it is
 not possible to perceive interference processes and it is very
 difficult to specify the  mechanism of magnetoresistance. The aim of our
 work is to investigate the influence of the interference on the
 spin-dependent transport through the SWCNT in the regime of strong coupling.

\section{Description of the model and calculation of the electronic current}

We consider the SWCNTs of the armchair type, which has 5 benzene
rings in the circumference. It is well known that this type CNT
has metallic electronic structure with two conducting channels.
The first and the last row of carbon atoms are connected with the
 electrodes, which are treated as ideal reservoirs.
\begin{figure}[!ht]
\center
\epsfxsize=6cm \epsfysize=3.5cm \epsfbox{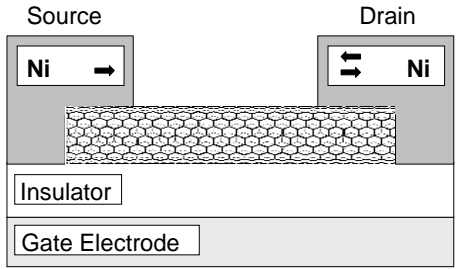} \caption{A
scheme of the system considered: A single walled carbon nanotube
of the armchair type (5,5) is attached to ferromagnetic electrodes
(e.g. Ni or Co). The relative polarization in both electrodes can be changed from the parallel to the antiparallel
orientation. The gate electrode voltage changes a relative energy of
incident electrons transmitted through the nanotube.}
\end{figure}
The system is described within the tight binding approach, in
which the hopping integrals are assumed $t_0=$ -2.5 eV for C--C bonds. The
hopping integral $t$ between the electrode and the carbon atom is
treated as a parameter. The current is calculated from the time
evolution of the electron number
$N_L=\sum_{i,k,\sigma}c^+_{kiL,\sigma}c_{kiL,\sigma}$ in the left
electrode
\begin{equation}
J=-e\left<\frac{{\rm d} N_L}{{\rm d}t}\right>=\frac{e}{\hbar}\;t
\sum_{i,k,\sigma}\int\frac{{\rm
d}\omega}{2\pi}\left[G^<_{1i\sigma,kiL\sigma}(\omega)+c.c.\right].
\end{equation}
Here, $G^<_{1i\sigma,kiL\sigma}(\omega)$ is the lesser Green
function connecting the carbons form the first row of the nanotube
with the left electrode. We sum over all incoming channels $i=$
1,...,10, wave vectors $k$ and spins $\sigma=\pm$ 1/2. The
non-equilibrium Green functions are determined from the Dyson
equation and the bare Green functions in the electrodes are taken
as $g_{\alpha\sigma}=i\pi\rho_{\alpha\sigma}$, where
$\rho_{L\sigma}$ and $\rho_{R\sigma}$ are densities of states for
electrons with the spin $\sigma$ at the Fermi energy  in the left
($\alpha=L$) and right ($\alpha=R$) electrode, respectively. Finally, we obtain the
formula
\begin{equation}
J=\frac{2e}{h}\sum_\sigma\frac{4}{\pi^2\rho_{L\sigma}\rho_{R\sigma}}\int{\rm
d}\omega[f_L(\omega)-f_R(\omega)]\sum_{i,j}|G^r_{Li,Rj}(\omega)|^2,
\end{equation}
where $G^r_{Li,Rj}$ is the retarded Green function connecting the
channels in the electrodes and $f_{\alpha}(\omega)$ is the
Fermi distribution function for electrons in the
$\alpha$-electrode.

\section{Conductance for the system with paramagnetic and ferromagnetic
 electrodes}

Fig. 2 presents the conductance
${\mathcal{G}}=\frac{dJ}{dV}\Bigl|_{V\to 0}$ calculated for the
SWCNT of the length of 90 atomic layers weakly and strongly connected to
the gold electrode (dashed and solid curve, respectively). For the
weak coupling ${\mathcal{G}}$ shows sharp resonant peaks at
energies $E_{ns}$ corresponding to standing electron waves in the
CNT. There are two branches ($s = 1,2$) of the dispersion
curves, and therefore
${\mathcal{G}}=\sum_{s,\sigma}{\mathcal{G}}_{s,\sigma}$ is a
superposition of the conductances for each channel $\{s,\sigma\}$.
For a resonant transmission (at $E_{ns}$)
${\mathcal{G}}_{s,\sigma}$ reaches its maximal value $e^2/h$. If
eigenvalues $E_{n1}$ and $E_{n2}$ are close to each other the
value of the conductance is larger than $2e^2/h$ (see outer peaks
on the left and the right hand side of the dashed curve in Fig. 2).
Next, we increase the coupling to the value $t=$ 1.3 eV, which
gives ${\mathcal{G}}$ close to the experimental data
3$e^2/h$~\cite{3}. For this case the resonant peaks are much
broader. If the peaks are close to each other, they merge together
leading to a single peak with the maximal value 4$e^2/h$. However,
the peaks can occur far from each other and then the effective
conductance plot looks like at the center of the Fig.2 (close to
$E=0$). The similar effect was observed experimentally \cite{3}.
Our approach takes into account multiple scattering on the
\begin{figure}[!t]
\center
\epsfxsize=10cm
\epsfysize=5cm
\epsfbox{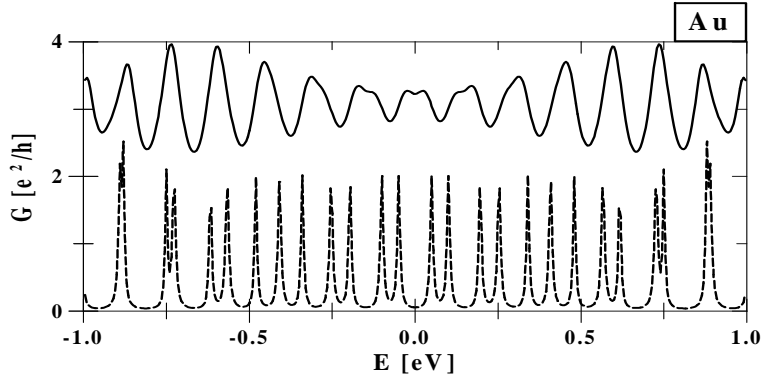}
\caption{Conductance ${\mathcal{G}}$ vs. incident electron energy
$E$ for the gold electrodes calculated for $T=0$. The solid curve
corresponds to the strong coupling  $t=$ 1.3 eV, and the dashed
curve to the weak coupling with  $t=$ 0.4 eV. The total density of
states at the Fermi energy is taken as $\rho_L=\rho_R=$ 0.294
states/eV \cite{seminario}.}
\end{figure}
contacts and interference processes. The electronic waves for both
conducting channels corresponding to the dispersion curves $s=1,2$ are of different symmetry and a transfer
matrix between them should be very small. In our opinion,
interference between the channels is irrelevant. It is in
contrast to \cite{3}, where a main role was assigned to the
Fabry-Perot interference.

\begin{figure}[!t]
\center
\epsfxsize=10cm
\epsfysize=9cm
\epsfbox{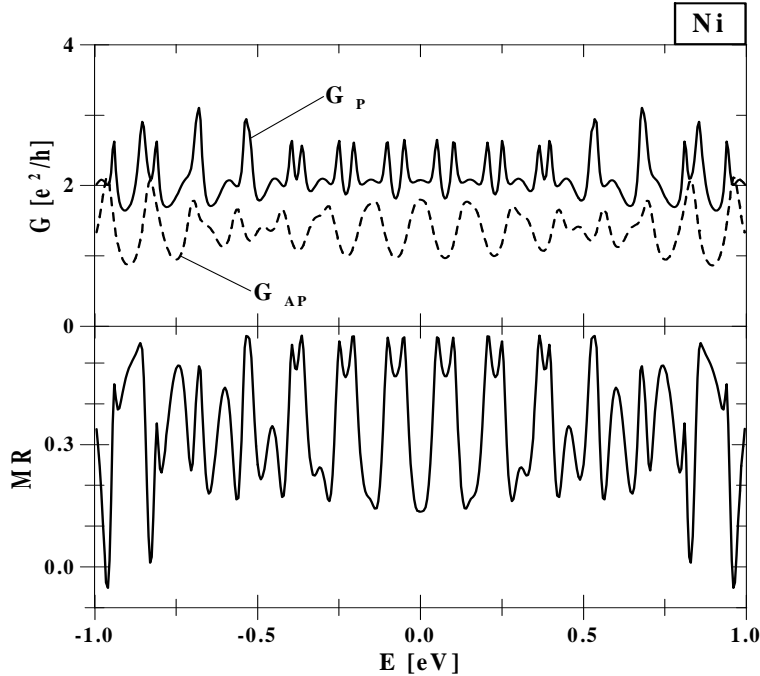}
\caption{The
conductance (a) ${\mathcal{G}}_P$ for the parallel (solid curve)
and ${\mathcal{G}}_{AP}$ the antiparallel configuration (dashed
curve) of the magnetization in the Ni electrodes, the
magnetoresistance (b) plotted as a function of the energy $E$ of
incident electrons. The parameters used are $t=$ 1.9 eV, $\rho_+=$
0.1897 state/eV and $\rho_-=$ 1.7261 states/eV.}
\end{figure}

\begin{figure}[!ht]
\center
\epsfxsize=10cm
\epsfysize=9cm
\epsfbox{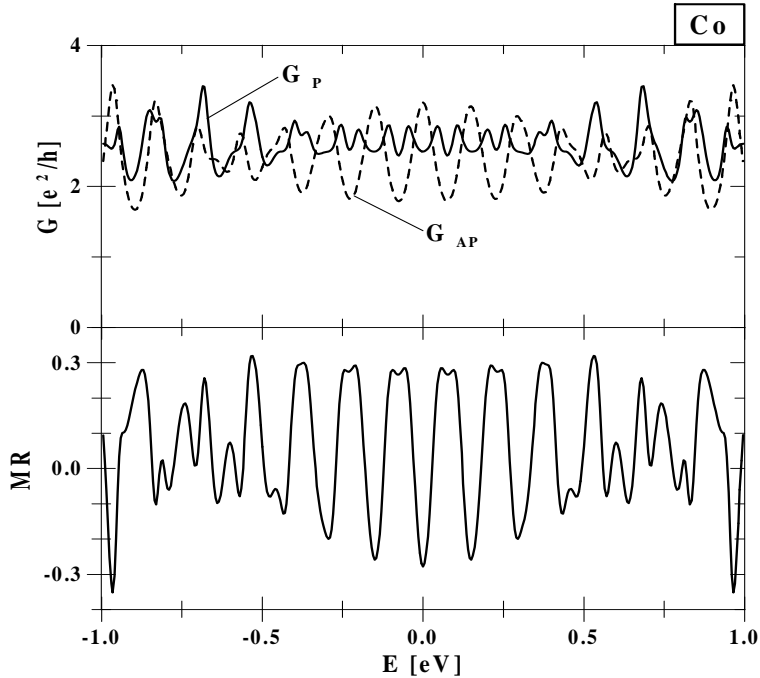}
\caption{The
conductance (a) ${\mathcal{G}}_P$ for the parallel (solid curve)
and ${\mathcal{G}}_{AP}$ the antiparallel configuration (dashed
curve) of the magnetization in the Co electrodes, the
magnetoresistance (b) plotted as a function of the energy $E$ of
incident electrons. The parameters used are $t=$ 1.9 eV, $\rho_+=$
0.1740 state/eV and $\rho_-=$ 0.7349 states/eV.}
\end{figure}

Next, we study transport through SWCNT attached to ferromagnetic
electrodes. The density of states were determined by the  band
structure calculations performed using the tight binding version
of the linear muffin-tin orbital method in the atomic sphere
approximation \cite{szajek}. The results are presented in Fig.3
for the system with the Ni electrodes. The conducting channels for
opposite spin orientations are different. Therefore,
${\mathcal{G}}$ is a superposition of four
${\mathcal{G}}_{s,\sigma}$ corresponding to the four different
channels $\{s,\sigma\}$. Since in ferromagnetic Ni $\rho_+=
0.1897$ states/eV for electrons with the up spin is much lower than
$\rho_-=$ 1.7261 states/eV for minority electrons with the down spin, so in the parallel
configuration ($P$) the conductance
${\mathcal{G}}_{+}=\sum_s{\mathcal{G}}_{s+}$ shows broad
peaks, which reach the value $2e^2/h$, while
${\mathcal{G}}_{-}$ shows sharp peaks and their maximal value
is $1e^2/h$. The composition of all channels results in
${\mathcal{G}}_P$ with sharp peaks (solid curve in Fig. 3a). The
situation is different for the antiparallel configuration ($AP$)
${\mathcal{G}}_{AP}$ (see the dashed curve in Fig. 3a), where the
peaks are relatively broad. The magnetoresistance
$MR=({\mathcal{G}}_P-{\mathcal{G}}_{AP})/{\mathcal{G}}_P$ is given
by the relative difference of the conductance in the parallel
and the antiparallel configuration. The results (presented in
Fig. 3b) show large changes of the $MR$ and that the maximal
value can be very large (max$[MR] >60\%$).

An interesting case is the system with the cobalt electrodes. The
density of states is $\rho_+=$ 0.1740 states/eV for the spin
$\sigma=+1/2$ (close to the value in Ni), but  $\rho_-=$ 0.7349
states/eV is much smaller. The peaks of the conductance
${\mathcal{G}}_{-}$ are now much broader. The total
conductance ${\mathcal{G}}_{P}$ and ${\mathcal{G}}_{AP}$ for both
orientations of polarization are shown in Fig.4a.
${\mathcal{G}}_{AP}$ is higher and shows larger oscillations than
for Ni (compare the dashed curves in Fig.3a and 4a). The value of
the magnetoresistance (shown in Fig.4b) is smaller, but exhibits
large oscillations. It is interesting that in this case $MR$ can
also be negative.

\section{Conclusions}

Our studies of coherent electronic transport through the SWCNT
showed that the conductance is a composition of contributions from
four channels. Although multiple scattering and interference
processes were included in this approach, we could not
observe any feature of destructive interference.  In our opinion,
the conducting channels correspond to waves of different symmetry
and matrix elements between them are small. We also
considered the spin-dependent transport in the system with the
ferromagnetic Ni and Co electrodes. Due to a resonant nature of
the electronic transport the magnetoresistance shows large
oscillations and achieves large values (max $[MR]\approx$ 60$\%$
and 30$\%$ for Ni and Co, respectively). Since the density of
states for minority electrons (with $\sigma=-1/2$) are different in
Ni and Co, the magnetoresistance is different in both cases.
For the Co electrodes $MR$ can even change the sign and can be
either positive or negative depending on the gate voltage applied
to the carbon nanotube. We believe that in near future one can
produce systems with coherent contacts between ferromagnetic
electrodes and the SWCNT, and that our theoretical predictions
will be verified.

\vspace*{0.25cm} \baselineskip=10pt{  \noindent \it We would like
to thank Dr. Andrzej Szajek for ab-initio calculations of the
density of states for ferromagnetic metals. The work was supported
by the Committee for Scientific Research (KBN) under Grant No.~2
P03B 087 19.}

\end{document}